\documentclass[compsoc,conference,a4paper,10pt,times]{IEEEtran}
\IEEEoverridecommandlockouts
\usepackage{cite}
\usepackage{amsmath,amssymb,amsfonts}
\usepackage{algorithmic}
\usepackage{graphicx}
\usepackage{textcomp}
\usepackage{bmpsize}
\usepackage{xcolor}
\usepackage{lipsum}

\def\BibTeX{{\rm B\kern-.05em{\sc i\kern-.025em b}\kern-.08em
		T\kern-.1667em\lower.7ex\hbox{E}\kern-.125emX}}

\usepackage{wasysym}

\usepackage{threeparttable}
\usepackage{progressbar}

\usepackage[inline]{enumitem}

\usepackage{booktabs}

\setenumerate{label=(\roman*), leftmargin=2em}

\newcommand{\inlineheading}[1]{\vspace{0.5em}\noindent\textbf{{#1}.}}

\newcommand{\censorhidden}[1]{}
\newcommand{\censorchange}[2]{#2}
\newcommand{\censor}[1]{\textit{\textless withheld during blind review \textgreater}}

\renewcommand{\censorhidden}[1]{#1}
\renewcommand{\censorchange}[2]{#1}
\renewcommand{\censor}[1]{#1}

\usepackage{soul}
\usepackage{xcolor}

\definecolor{hlcolor}{RGB}{214, 239, 255}\sethlcolor{hlcolor}%
\newcommand{\highlight}[1]{\ctext[RGB]{214, 239, 255}{#1}}
\newcommand{\highlightimage}[1]{
	\setlength{\fboxsep}{0pt}%
	\setlength{\fboxrule}{1pt}%
	\fcolorbox{hlcolor}{hlcolor}{
	\hspace{-1.5mm}#1}
}

\renewcommand{\highlight}[1]{#1}
\renewcommand{\highlightimage}[1]{#1}

\usepackage{hyperref}

\usepackage{textpos}

\begin{document}
	
	\title{Privacy Considerations for Risk-Based Authentication Systems\thanks{This research was supported by the research training group ``Human Centered Systems Security'' (NERD.NRW) sponsored by the state of North Rhine-Westphalia.}}

	\author{\IEEEauthorblockN{
	Stephan Wiefling\IEEEauthorrefmark{1}, 
	Jan Tolsdorf, 
	and Luigi Lo Iacono}
		\IEEEauthorblockA{H-BRS University of Applied Sciences, Sankt Augustin, Germany\\\IEEEauthorrefmark{1}Ruhr University Bochum, Bochum, Germany\\
		\{stephan.wiefling,jan.tolsdorf,luigi.lo\_iacono\}@h-brs.de}
	}
	
	\maketitle
	\begin{textblock*}{\textwidth}(0cm,-7cm)
		\centering
		2021 IEEE European Symposium on Security and Privacy Workshops (EuroS\&PW)
	\end{textblock*}
	\vspace{-1.3em}%
	\begin{textblock*}{0.5\textwidth}(0cm,20.6cm)
		\noindent
		{\footnotesize \textcopyright~2022 Stephan Wiefling, Jan Tolsdorf, Luigi Lo Iacono}\\
		{\footnotesize Open Access version of a paper published at IWPE '21.\\DOI: \href{https://doi.org/10.1109/EuroSPW54576.2021.00040}{10.1109/EuroSPW54576.2021.00040}}\\
	\end{textblock*}

    \pagestyle{plain}

	\begin{abstract}
Risk-based authentication (RBA) extends authentication mechanisms to make them more robust against account takeover attacks, such as those using stolen passwords. RBA is recommended by NIST and NCSC to strengthen password-based authentication, and is already used by major online services. Also, users consider RBA to be more usable than two-factor authentication and just as secure. However, users currently obtain RBA's high security and usability benefits at the cost of exposing potentially sensitive personal data (e.g., IP address or browser information). This conflicts with user privacy and requires to consider user rights regarding the processing of personal data.

We outline potential privacy challenges regarding different attacker models and propose improvements to balance privacy in RBA systems. To estimate the properties of the privacy-preserving RBA enhancements in practical environments, we evaluated a subset of them with long-term data from 780 users of a real-world online service.
Our results show the potential to increase privacy in RBA solutions. However, it is limited to certain parameters that should guide RBA design to protect privacy. We outline research directions that need to be considered to achieve a widespread adoption of privacy preserving RBA with high user acceptance. 	\end{abstract}
	
	\begin{IEEEkeywords}
		Password, Risk-based Authentication, Usable Security and Privacy, Big Data Analysis
	\end{IEEEkeywords}

\section{Introduction}

Passwords are still predominant for authentication with online services~\cite{quermann_state_2018}, although new threats are constantly emerging. Credential stuffing and password spraying attacks~\cite{haber_attack_2020} use leaked login credentials (username and password) sourced from data breaches, and try them in some way on (other) online services. These attacks are very popular today~\cite{akamai_loyalty_2020} since attackers can automate them with little effort%
.
Major online services responded to this threat with implementing risk-based authentication (RBA)~\cite{wiefling_is_2019}, aiming to strengthen password-based authentication with little impact on the user.

\subsubsection*{Risk-Based Authentication (RBA)}
RBA determines whether a login attempt is a legitimate one or an account takeover attempt.
To do so, RBA monitors additional features when users submit their login credentials. Popular features range from network (e.g., IP~address), device (e.g., smartphone model and operating system), or client (e.g., browser vendor and version), to (behavioral) biometric information (e.g., login time)~\cite{wiefling_whats_2021,wiefling_is_2019}. Based on the feature values and those of previous logins, RBA calculates a risk score. An access threshold typically classifies the score into low, medium, and high risk~\cite{freeman_who_2016,hurkala_architecture_2014,molloy_risk-based_2012}. On a low risk (e.g., usual device and location), the RBA system grants access with no further intervention. On a medium or higher risk (e.g., unusual device and location), RBA requests additional information from the user, e.g., verifying the email address. After providing the correct proof, access is granted.%

RBA is considered a scalable interim solution when passwords cannot simply be replaced by more secure authentication methods in many cases~\cite{wiefling_more_2020,wiefling_whats_2021}. The National Institute of Standards and Technology (NIST, USA) and National Cyber Security Centre (NCSC, UK) recommend RBA to mitigate attacks involving stolen passwords~\cite{grassi_digital_2017,national_cyber_security_centre_cloud_2018}. Beyond that, users found RBA more usable than equivalent two-factor authentication (2FA) variants and comparably secure~\cite{wiefling_more_2020}. Also, in contrast to 2FA, RBA both offers good security and rarely requests additional authentication in practice~\cite{wiefling_whats_2021}, reducing the burden on users.

\subsubsection*{Research Questions}

However, users obtain the security and usability gain of RBA at the cost of disclosing more potentially sensitive data with a personal reference, such as IP addresses and browser identifiers. Therefore, user privacy is at risk when RBA databases are forwarded or breached, as additional data besides usernames would potentially allow to identify individuals.

More and more data protection laws aim to protect users from massive data collection by online services. Considering that, we wondered whether and to what extent the integration of RBA systems complies with the principles of modern data protection. We also wondered which trade-offs are possible to balance security and privacy goals.

To further investigate RBA's privacy aspects, we formulated the following research questions:

\newlist{RQLIST}{enumerate}{1}
\setlist[RQLIST]{label=\bfseries RQ\arabic*:, leftmargin=2.7em, parsep=0em}

\newlist{RQ2LIST}{enumerate}{2}
\setlist[RQ2LIST]{label=\bgroup\bfseries \alph*)\egroup,leftmargin=1.3em, parsep=0em}

\begin{RQLIST}
	\item \begin{RQ2LIST}
	    \item In what ways can RBA features %
	    be stored to increase the user privacy?
	    \item How can RBA features %
	    be stored to protect user privacy in terms of data breaches?
        \end{RQ2LIST}
	\item To what extent can a RBA feature maintain good security while preserving privacy in practice?
\end{RQLIST}

\subsubsection*{Contributions}
We propose and discuss five privacy enhancements that can be used by RBA models used by the majority of deployments found in practice. To estimate their usefulness in practice, we evaluated a subset of these enhancements on a RBA feature that is highly relevant in terms of security and privacy, i.e., the IP address. We evaluated with a data set containing the login history of 780 users on a real-world online service for over 1.8 years%
. %

Our results show for the first time that it is possible to increase feature privacy while maintaining RBA's security and usability properties. However, increasing privacy is limited to certain conditions that need to be considered while designing the RBA system. We also identified future challenges and research directions that might arise with a widespread RBA adoption in the future.

The results support service owners to provide data protection compliant RBA solutions. They assist developers in designing RBA implementations with increased privacy. Researchers gain insights on how RBA can become more privacy friendly%
, and further research directions.

\section{Background}\label{sec:background}
In the following section, we provide a brief introduction to RBA and explain how the use of RBA correlates with the several privacy principles defined by industry standards and legislation.

\subsection{RBA Model} \label{subsec:rba-model}

Since RBA is not a standardized procedure, multiple solutions %
exist in practice. We focus on the implementation by Freeman et al.~\cite{freeman_who_2016}, since it performed best in a \censorchange{previous study}{study of Wiefling et al.}~\cite{wiefling_whats_2021}. Also, this RBA model is known to be widely used, e.g., by popular online services like Amazon, Google, and LinkedIn~\cite{wiefling_is_2019,wiefling_whats_2021}.

The model calculates the risk score $S$ for a user $u$ and a set of feature values $(FV^1,..., FV^d)$ with $d$ features as:
\begin{equation}
    S_{u}(FV) = \left( \prod_{k=1}^{d} \frac{p(FV^k)%
    }{p(FV^k | u, legit)%
    } \right) \frac{p(u | attack)}{p(u | legit)}
\end{equation}
$S$ has the probabilities $p(FV^k)$ that a feature value appears in the global login history of all users, and \linebreak $p(FV^k | u, legit)$ that a legitimate user has this feature value in its own login history. 
The probability $p(u | attack)$ describes how likely the user is being attacked, and $p(u | legit)$ describes how likely the legitimate user is logging in.

\subsection{Regulatory Foundations}

\highlight{In the past few years, the introduction of new data protection laws, such as the General Data Protection Regulation (GDPR)}~\cite{european_union_gdpr_2016} \highlight{and the California Consumer Privacy Act (CCPA)}~\cite{california_ccpa_2018}, \highlight{dramatically changed the way online services (i.e., data controllers) process their users' data.
Formerly loose recommendations on handling user data have been replaced by clear and binding data protection principles, which data controllers must adhere to. However, the details and scope of the principles vary between jurisdictions. For internationally operating data controllers, this poses the problem that their data processing operations must be designed to be compatible with different requirements. Fortunately, the privacy framework specified in ISO 29100:2011}~\cite{iso_isoiec_2011} \highlight{already compiles an intersection of privacy principles from data protection laws worldwide. Thus, it provides data controllers a solid basis for designing legally compliant data processing operations that can be tailored to the details of different jurisdictions. We outline the requirements for the design of RBA systems based on the privacy principles defined in ISO 29100:2011, aiming at compatibility with different jurisdictions.}

\inlineheading{Applicability of Privacy \highlight{Principles}}
Generally speaking, \highlight{the privacy principles defined in established privacy laws and frameworks aim} to protect the privacy of individuals. Thus, they only apply to data with a personal reference. Such data are called, e.g., \emph{``personal data''} (GDPR \cite{european_union_gdpr_2016}), \emph{``personal information''} (CCPA \cite{california_ccpa_2018}), or \emph{``personally identifiable information''} (PII) (ISO \cite{iso_isoiec_2011}).
The definitions are very similar and usually refer to \emph{``any information that (a) can be used to identify [an individual] to whom such information relates, or (b) is or might be directly or indirectly linked to [an individual]''} \cite{iso_isoiec_2011}. 

\highlight{The data processed by RBA certainly fall within this definition,} since 
implementations 
rely on features that already serve as (unique) identifiers by themselves (e.g., IP address)~\cite{wiefling_is_2019}. \highlight{Also, the risk score calculated by RBA represents an identifier by itself, as it constitutes a set of characteristics that uniquely identifies an individual.} %
Therefore, RBA has to comply with \highlight{ISO~29100:2011's} privacy principles discussed below.

\inlineheading{Consent and Choice} 
\highlight{In general, data controllers must ensure the lawfulness of data processing. While most jurisdictions recognize user consent as a lawful basis, applicable laws may allow processing without consent. Depending on the assets associated with a user account, data controllers may argue that RBA use is required to comply with the obligation to implement appropriate technical safeguards against unauthorized access. Nonetheless, to ensure compliance, providers should design RBA mechanisms with consent in mind and provide their users with clear and easy-to-understand explanations.%
}

\inlineheading{Collection Limitation and Data Minimization} 
Data controllers must limit the PII collection and processing to what is necessary for the specified purposes. \highlight{RBA feature sets should therefore be reviewed for suitability with redundant or inappropriate features removed}~\cite{wiefling_whats_2021}.
\highlight{This includes considering using pseudonymized data for RBA and disposing of the feature values when they are no longer useful for the purpose of RBA.
In practice, this creates the challenge to not reduce a risk score's reliability.}

\inlineheading{Use, Retention, and Disclosure Limitation}
The data processing must be limited to purposes specified by the data controller, and data must not be disclosed to recipients other than specified. RBA should ensure that features cannot be used for purposes other than the calculation of risk scores. Moreover, after a feature value becomes outdated, it should be securely destroyed or anonymized. We would point out that privacy laws do not apply to anonymized data and could therefore serve data controllers for developing and testing purposes beyond the retention period specified in their privacy statements.

\inlineheading{Accuracy and Quality} Data controllers must ensure that the processed data are accurate and of quality. This is not only due to their own business interests, but also because data subjects have a right to expect their data being correct. This directly affects RBA, since it has the power to deny a significant benefit to users (i.e., access to their user account) with potentially significant harm. Data controllers must hence ensure by appropriate means that the stored feature values are correct and valid.

\inlineheading{Individual Participation and Access} Data controllers must allow data subjects to access and review their PII. For RBA, this means that users should be allowed to be provided with a copy of the feature values used.

\inlineheading{Information Security}
Data controllers are obliged to protect PII with appropriate controls at the operational, functional, and strategic level against risks. These %
include, but are not limited to, risks associated with unauthorized access or processing and denial of service. 
Privacy laws demand extensive protections in this regard,
\emph{``taking into account the state of the art, the costs of implementation and the nature, scope, context and purposes of processing as well as the risk of varying likelihood and severity for the rights and freedoms of natural persons''} (Art. 32 (1) GDPR).
\highlight{Since RBA risk scores do not necessarily rely on evaluating plain text feature values}~\cite{wiefling_whats_2021}\highlight{, the collected data should be stored in an appropriate pseudonymized, masked, truncated, or encrypted form, depending on the RBA implementation.}
Moreover, data controllers \highlight{should implement additional technical and organizational measures as needed,} and be able to ensure the integrity, availability, and resilience of RBA.

\inlineheading{Accountability and Privacy Compliance} Data controllers should inform data subjects about privacy-related policies, transfers of PII to other countries, and data breaches. Data controllers should also implement organizational measures to help them verify and demonstrate legal compliance. These include, but are not limited to, risk assessments and recovery procedures. RBA implementations should therefore consider the worth of RBA features to both attackers and data subjects, and the recovery from data breaches. This is crucial in order not to undermine the security of user accounts and their associated assets.

\section{Privacy Enhancements (RQ1)} \label{sec:privacy-considerations}

To comply with the privacy principles and derived data protection requirements, service owners should consider mechanisms to increase privacy in their RBA implementations.
In the following, we introduce threats and their mitigation to increase privacy properties of RBA features.

\subsection{Feature Sensitivity and Impact Level} \label{subsec:feature-sensitivity}

RBA feature sets always intend to distinguish attackers from legitimate users. In doing so, the features may contain sensitive PII. However, not only do users perceive such PII differently regarding their sensitivity~\cite{schomakers_internet_2019}. Their (unintended) disclosure could also have far-reaching negative consequences for user privacy.
Developers and providers should therefore determine the impact from a loss of confidentiality of the RBA feature values. Specifically, the following aspects need consideration \cite{mccallister_guide_2010}:

\inlineheading{Identifiability\highlight{ and Linkability}}
RBA feature sets should be evaluated regarding their ability to identify natural persons behind them.
In particular, RBA systems that rely on intrusive online tracking methods, such as browser fingerprinting, store %
sensitive browser-specific information that form a %
linked identifier. In the event of losing confidentiality,
the features would allow clear linkage between profiles at different online services, despite \highlight{users using different login credentials or pseudonyms}. Depending on the service, this could result in negative social or legal consequences for individuals. It could also enable more extensive and unintended activity tracking, and de-anonymizing information associated with user accounts%
.
Previous work found that powerful RBA feature sets do not require to uniquely identify users when focusing on the detection of account takeover attempts~\cite{wiefling_whats_2021}. %
Also, users are more willing to accept the processing of sensitive information when they are certain that it is anonymous and does not allow them to be identified~\cite{markos_new_2018,schomakers_all_2020}. Thus, the use of non-intrusive features may increase user trust in online services, too.

\inlineheading{Feature Values Sensitivity}
Aside from %
identifying individuals by RBA feature sets, the individual feature values may already contain sensitive PII. Sensitive PII in the scope of RBA may be feature values that are easily spoofable and can be misused to attack other online services in the event of a data breach. Sensitive PII may also refer to data perceived as sensitive by online users.  
For example, 
the most important feature of current RBA methods, namely the IP address~\cite{wiefling_whats_2021,wiefling_is_2019,freeman_who_2016,steinegger_risk-based_2016,hurkala_architecture_2014}, is perceived as highly sensitive by online users of diverse cultural backgrounds~\cite{markos_information_2017,schomakers_internet_2019,almotairi_perception_2020}. Since users are generally less willing to share data with increased sensitivity, RBA feature sets should limit the use of sensitive data if possible, in order to meet user interests.

\subsection{Threats}

RBA features may contain personal sensitive data, which has to be protected against attackers. To support online services in their protection efforts, we introduce \highlight{three privacy threat types}. We based the threats on those found in literature and our own observations in practice.

\inlineheading{Data Misuse} \highlight{Online services could misuse their own RBA feature data for unintended purposes, such as user tracking, profiling, or advertising} ~\cite{bonneau_privacy_2014}. \highlight{This type of misuse previously happened with phone numbers stored for 2FA purposes}~\cite{venkatadri_investigating_2018}. \highlight{While users have to trust online services to not misuse their data, responsible online services should also take precautions to minimize chances for miuse scenarios or unintended processing, e.g., by internal misconduct or after the company changed the ownership.}

\inlineheading{Data Forwarding} Online services can be requested or forced to hand out stored feature data, e.g., to state actors, advertising networks, or other third parties. Especially IP addresses are commonly requested%
~\cite{europol_sirius_2019}. When such data are forwarded to third parties, the users' privacy is breached. For instance, the IP address could be used to reveal the user's geolocation or even their identity.%

\inlineheading{Data Breach} Attackers %
obtained the database containing the feature values%
, e.g., by hacking the online service%
.
As a result, online services lost control over their data. Attackers can try to re-identify users based on the feature values, e.g., by combining them with other data sets. They can further try to reproduce the feature values and try account takeover attacks on a large scale, similar to credential stuffing. On success, they could access sensitive user data stored on the online service, e.g., private messages. %

\subsection{Mitigation}

Online services can implement several measures to mitigate the outlined privacy threats. We propose five %
measures that are based on methods found in related research fields, as well as privacy regulations and our own observations with the selected RBA model (see Section~\ref{sec:background}).
Based on the introduced RBA model, we considered all feature values as categorical data, in order to calculate the probabilities. When this condition is met, the proposed measures are also applicable to other RBA models~\cite{wiefling_whats_2021}%
.%

As an example for practical solutions, we describe how the considerations can be applied to the \emph{IP address} feature, with regard to the IPv4 address. We chose this feature since it is both considered the most important RBA feature in terms of security to date and sensitive re-linkable data in terms of privacy (see Section~\ref{subsec:feature-sensitivity}).

\subsubsection{Aggregating}

The RBA model only depends on feature value frequencies%
. To minimize data and limit misuse~\cite{iso_isoiec_2011}, we can aggregate or reorder feature data in the login history without affecting the results. The data set would then reveal how often a feature combination occurred, but not its chronological order. Removing this %
information can %
mitigate re-identification in login sequences. 

\subsubsection{Hashing}

A cryptographic hash function, such as SHA-256, transforms a data input of arbitrary value to an output of fixed length%
. As inverting a hash function is not possible in theory, attackers need to recalculate all possible hashing values to restore the input values~\cite{%
llewellyn-jones_cracking_2017}. Assuming that the hashes are practically collision-free, using hashed feature values will produce the same RBA results as with the original values. This is the case, because the feature values are only transformed into a different representation%
. Therefore, this could be a solution to protect feature data in terms of the outlined threats. %

However, the IPv4 address has 32 bit limited input values, where some addresses have a specific semantic and purpose, and cannot be assigned to devices. Thus, %
attackers can simply hash all $2^{32} -1$ values to restore the correct IP address. To counteract this problem, we can append a large random string (salt) to the input value:
\begin{equation}
    H(192.168.1.166\ ||\ salt) = 243916...aad132
\end{equation}
Attackers need to guess the salt correctly, which is high effort when the salt is large. Thus, this mitigation strategy increases the required guessing time for each feature value. Taking it a step further, we can even hash the results multiple times to increase the computation time%
:
\begin{equation}
    H(H(...H(192.168.1.166\ ||\ salt))) = [hash]
\end{equation}
This is similar to key derivation strategies used in password databases~\cite{moriarty_pkcs_2017}. However, we can only use a global salt for all database entries, as RBA mechanisms need to be able to identify identical feature values across users in the database. By increasing the computational cost, attackers cannot scale attacks as they would have with the unhashed feature values.%

\subsubsection{Truncation}

A more destructive approach to increase privacy for RBA features is to change or remove details from their data values. This can reduce the number of records with unique quasi identifiers. Since the feature data then becomes less useful for other use cases like tracking or re-identification, we consider it a measure to mitigate the privacy threats. Regarding the IP address, we could set the last bits to zero. For truncating the last eight bits, for example, this would result in:
\begin{equation}
    Truncate(192.168.1.166, 8\ Bit) = 192.168.1.0
\end{equation}
This mechanism is known from IP address anonymization strategies~\cite{chew_privacy_2019,burkhart_risk-utility_2008}. However, we can also apply it on other features, e.g., reducing timing precision or coarse-graining browser version number in the user agent string~\cite{pugliese_long-term_2020}. Since we remove information that could potentially identify an individual, e.g., the device's internet connection, this can potentially increase privacy. However, this can also influence the RBA results, as there are fewer feature values for attackers to guess. %

\subsubsection{K-Anonymity}

The k-anonymity privacy concept~\cite{sweeney_k-anonymity_2002} ensures that at least $k$ entries in a data set have the same quasi identifier values. If attackers obtained the data set and know a victim's IP address, they would not be able to distinguish the person from $k$ other users.
This makes it an effective countermeasure against re-identification in case of data forwarding and data breaches.

To achieve k-anonymity for RBA, at least $k$~users need to have the same feature value. To ensure this, we added potentially missing entries to the RBA login history after each successful login. We added these entries to random users to only affect the global login history probabilities in order to keep a high security level.
We created these users just for this purpose. To retain the global data set properties, the user count increased gradually to have the same mean number of login attempts per user.

\subsubsection{Login History Minimization}

Another approach is to limit the login history, in terms of the amount of features and entries, for a number of entries or a constant time period~\cite{iso_isoiec_2011}. A study already showed that few entries are sufficient to achieve a high RBA protection~\cite{wiefling_whats_2021}. In so doing, we mitigate tracking users for an extended period of time. However, this can affect the RBA performance based on the usage pattern of the corresponding online service. Especially when it is a less-than-monthly-use online service, we assume that features need to be stored for a longer period than for daily use websites to achieve a comparable RBA performance.

\section{Case Study Evaluation (RQ2)} \label{sec:case-study-evaluation}

Aggregating and hashing, when collision-free, does not affect the RBA results, as they only change the data representation for the RBA model. The other approaches, however, potentially could. To assess their impact on RBA behavior in practice, we studied truncation and k-anonymity using real-world login data. The properties and limited size of our data set did not allow to reliably test the login history minimization approach, so we left it for future work. Nevertheless, we outlined this relevant privacy consideration for the sake of completeness. We used the IP address feature as in the other examples.

\subsection{Data Set}

For the evaluation, we used \censorchange{our}{a} long-term RBA data set\censorchange{}{ by Wiefling et al.}, including features of 780 users collected on a real-world online service~\cite{wiefling_whats_2021}. The online service collected the users' features after each successful login. The users signed in 9555 times in total between August 2018 to June 2020. They mostly logged in daily (44.3\%) or several times a week (39.2\%), with a mean of 12.25 times in total. To improve data quality and validity, \censorchange{we}{they} removed all users who noticed an illegitimate login in their account. The online service was an e-learning website, which students used to exercise for study courses and exams. As the users were mostly located in the same city, it is a very challenging data set for RBA. They could get similar IP addresses with higher probability. Therefore, it is important to evaluate how the RBA protection changes in such a challenging scenario.

\subsubsection{Legal and Ethical Considerations}

The study participants\censorchange{}{ of Wiefling et al.}~\cite{wiefling_whats_2021} signed a consent form agreeing to the data collection and use for study purposes. \censorchange{}{They also agreed to using the data in this analysis. }They were always able to view a copy of their data and delete it on request. The collected data were stored on encrypted hard drives and only the researchers had access to it.

We do not have a formal IRB process at our university. Still, we made sure to minimize potential harm by complying with the ethics code of \censorchange{the German Sociological Association (DGS)}{a nationwide sociological association} and the standards of good scientific practice of \censorchange{the German Research Foundation (DFG)}{a nationwide research funding organisation\footnote{\label{footnote:orgname-omitted}Organization names omitted during blind review}}. We also made sure to comply with the GDPR.

\subsubsection{Limitations}

Our results are limited to the data set and the users who participated in the study. They are limited to the population of a certain region of a certain country. They are not representative for large-scale online services, but show a typical use case scenario of a daily to weekly use website. As in similar studies, we can never fully exclude that intelligent attackers targeted the website. However, multiple countermeasures minimized the possibility that the website was infiltrated~\cite{wiefling_whats_2021}.

\subsection{Attacker Models}

We evaluated the privacy enhancements using three RBA attacker models found in related literature~\cite{wiefling_whats_2021,freeman_who_2016}. %
All attackers possess the login credentials of the target.%

\textbf{Naive attackers} try to log in from a random Internet Service Providers (ISP) from somewhere in the world. We simulated these attackers by using IP addresses sourced from real-world attacks on online services~\cite{firehol_all_2020}.

\textbf{VPN attackers} know the country of the victim. Therefore, we simulated these attackers with IP addresses from real-world attackers located in the victim's country~\cite{firehol_all_2020}.

\textbf{Targeted attackers} know the city, browser, and device of the victim. Therefore, they choose similar feature values, including similar ISPs. We simulated these attackers with our data set, with the unique feature combinations from all users except the victim. Since the IP addresses of our data set were in close proximity to each other, our simulated attacker was aware of these circumstances and chose them in a similar way.

\subsection{Methodology}

In order to test our privacy enhancements in terms of practical RBA solutions, we defined a set of desired properties. Our enhancements need to: \begin{enumerate*}[label=(\Alph*)]
    \item \textbf{Keep the percentage of blocked attackers}: The ability to block a high number of attackers should not decrease when using the privacy enhancements. This is necessary to keep the security properties of the RBA system.
    \item \textbf{Retain differentiation between legitimate users and attackers}: When applied, the risk score differences between legitimate users and attackers should only change within a very small range. Otherwise, the usability and security properties of the RBA system would decrease.
\end{enumerate*}

We outline the tests to evaluate the privacy enhancements below. Based on the method in Wiefling et al.~\cite{wiefling_whats_2021}, we reproduced the login behavior for attackers and legitimate users by replaying the user sessions. We integrated truncation and k-anonymity in the reproduction process, to test the countermeasures.

The RBA model used the IP address and user agent string as features, since this can be considered the RBA state of practice~\cite{wiefling_is_2019,wiefling_whats_2021}. We truncated the IP addresses in ranges from 0 to 24 bits, to observe the effects on the RBA performance. We assume that cutting more than 25 bits %
will not allow to reliably detect attackers. We also tested k-anonymity with the IP address feature until $k=6$. As US government agencies consider less than five entries to be sensitive~\cite{federal_committee_on_statistical_methodology_report_2005}, we chose to cover this threshold.

\subsubsection{Test A: Percentage of Blocked Attackers}

To compare the RBA performance regarding all three attacker models, we calculated the percentage of how many attackers would be blocked. We call this percentage the \emph{true positive rate} (TPR), as previous work did~\cite{wiefling_whats_2021,freeman_who_2016}. For a fair comparison, we observed how the TPR changed when aiming to block 99.5\% of attackers. We chose this TPR baseline since it showed good performance regarding usability and security properties in a previous study~\cite{wiefling_whats_2021}.

To ease comparison, we adjusted the TPR for each truncation or k-anonymity step $x_i$ as percentage differences to the baseline without modifications (relative TPR):
\begin{equation}
    TPR_{relative_{x_i}} = \frac{TPR_{x_i} - TPR_{baseline}}{TPR_{baseline}}
\end{equation}
Following that, $TPR_{relative_{x_i}} < 0.0$ means that the TPR decreased compared to the baseline.

\subsubsection{Test B: Risk Score Changes}

To determine the degree that attackers and legitimate users can be differentiated in the RBA model, we calculated the \emph{risk score relation} (RSR)~\cite{wiefling_whats_2021}. It is the relation between the mean risk scores for attackers and legitimate users:
\begin{equation}
    RSR_{basic} = \frac{mean\ attacker\ risk\ score}{mean\ legitimate\ risk\ score}
\end{equation}
To ease comparison, we normalized each RSR for every truncation or k-anonymity step $x_i$ as percentage differences to the baseline (relative RSR). The baseline is the IP address without modifications:
\begin{equation}
    RSR_{relative_{x_i}} = \frac{RSR_{basic_{x_i}} - RSR_{baseline}}{RSR_{baseline}}
\end{equation}
As a result, $RSR_{relative_{x_i}} < 0.0$ signals that attackers and legitimate users can no longer be distinguished as good as they were before introducing the privacy enhancing measures.

\subsubsection{Limit Extraction}

For each test, we defined the following thresholds to extract limits that do not degrade RBA performance to an acceptable extent.

(Test A) We require the RBA performance to remain constant. Thus, we selected the reasonable limit as the point at which the relative TPR decreases compared to the baseline, i.e., attackers cannot be blocked as good as before any more.
(Test B) Unlike tracking, RBA uses the feature information in addition to an already verified identifier, e.g., passwords. Thus, we consider it feasible to reduce the RSR slightly for the sake of privacy. Based on our observations%
, RSR changes below 0.01 can be tolerable for our case study evaluation. Thus, we chose the reasonable limit as the point at which the relative RSR is lower than 0.01.

\subsection{Results}

In the following, we present the results for all attacker models. We discuss the results after this section. We used a high performance computing cluster using more than 2000 cores for the evaluation. This was necessary since calculating the results with the simulated attackers was computationally intensive.

For statistical testing, we used Kruskal-Wallis tests for the omnibus cases and Dunn's multiple comparison test with Bonferroni correction for post-hoc analysis. We considered p-values less than 0.05 to be significant.

\subsubsection{Truncation}

\begin{figure}
    \centering
    \highlightimage{\includegraphics[width=\linewidth]{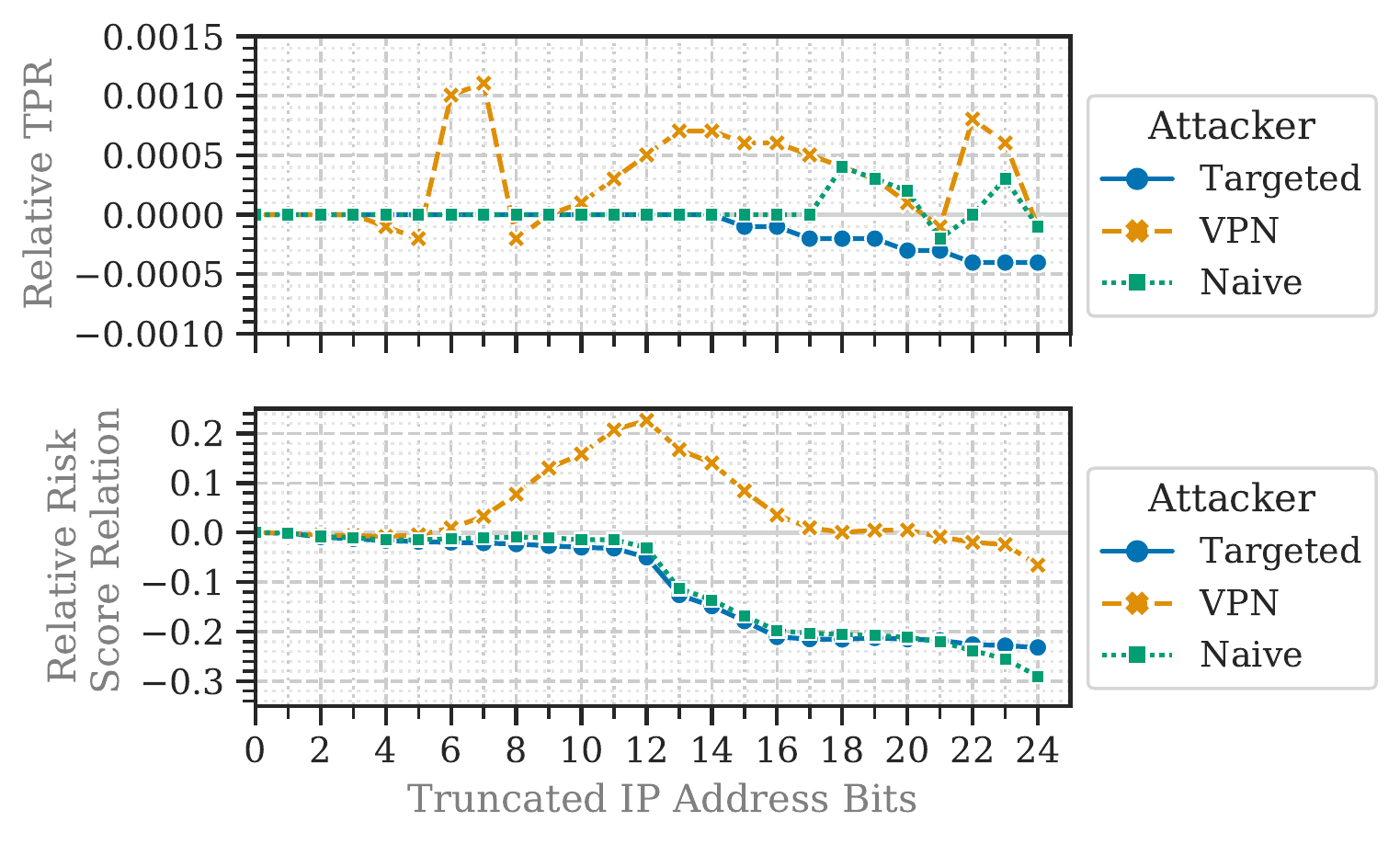}}
    \caption{Results for truncating the IP address. Top: Relative TPR (Test A). There were significant differences between targeted and both VPN and naive attackers. Bottom: Relative RSR (Test B). The differences between VPN and both targeted and naive attackers were significant.}
     \label{fig:relative-rsr-tpr}
\end{figure}

Figure~\ref{fig:relative-rsr-tpr} shows the %
truncation test results for all attackers. The TPR differences between the targeted attacker and both remaining attackers were significant (Targeted/Naive: p=0.0151, Targeted/VPN: p$<$0.0001). The TPRs exceeded the limit after 20 bits for naive, 3 bits for VPN, and 14 bits for targeted attackers.

Regarding the relative RSRs, there are significant differences between VPN and both remaining attackers (p$<$0.0001).
The RSRs exceeded the limit after 3 bits for naive, 21 bits for VPN, and 3 bits for targeted attackers.

Combining both results, the accepted truncation limits based on our criteria were 3 bits for all %
attacker models.

\subsubsection{K-Anonymity}

\begin{figure}
    \centering
    \highlightimage{\includegraphics[width=\linewidth]{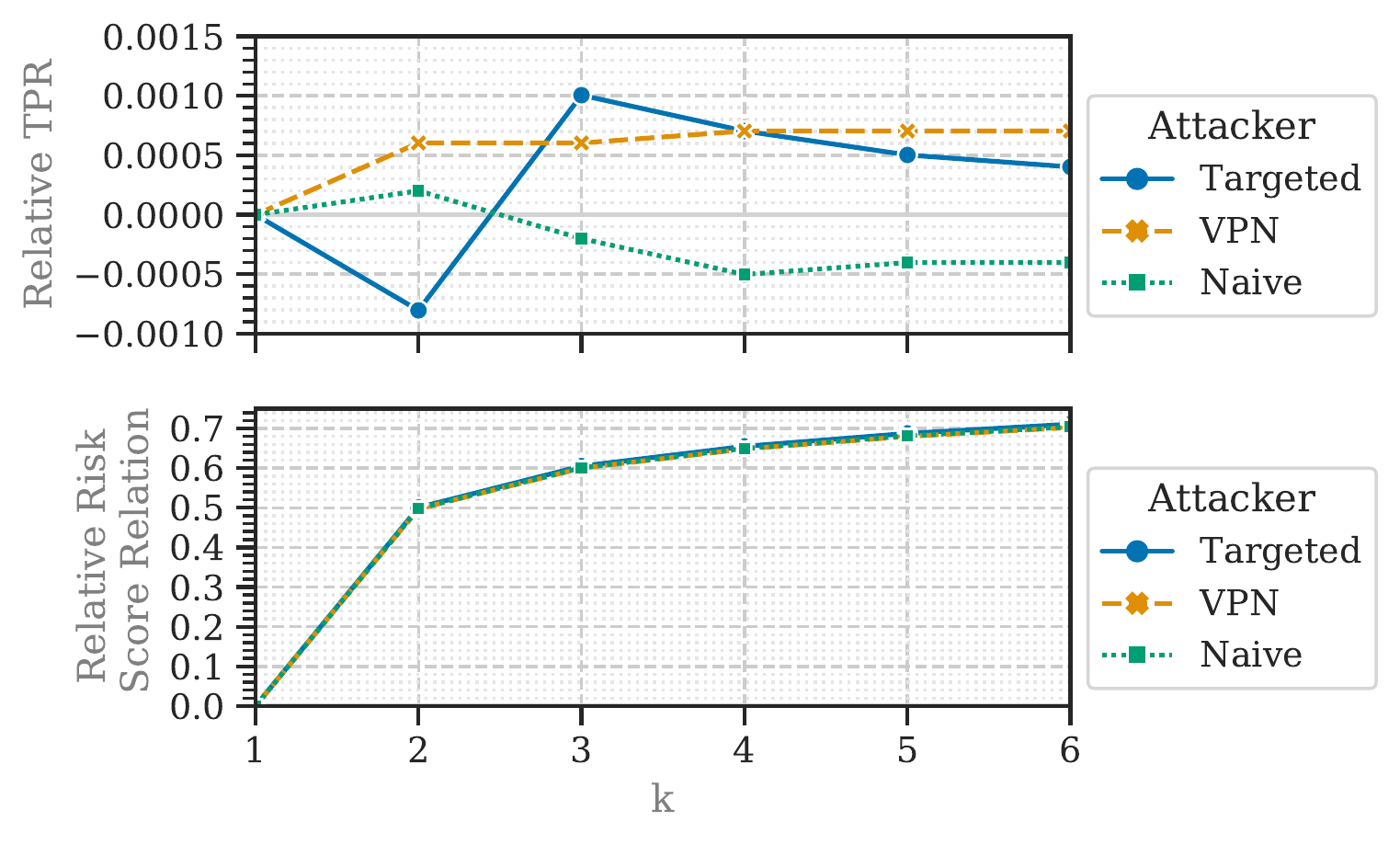}}
    \caption{Results for k-anonymity regarding the IP address. Top: Relative TPR (Test A). Differences between naive and VPN attackers were significant. Bottom: Relative RSR (Test B). There were no significant differences.}
    \label{fig:relative-rsr-tpr-k}
\end{figure}

Figure~\ref{fig:relative-rsr-tpr-k} shows the combined k-anonymity test results for the three attacker models. The relative TPR decreased after $k=1$ for targeted attackers, $k=2$ for naive attackers, and not at all for VPN attackers until at least $k=6$. There were significant TPR differences between naive and VPN attackers (p=0.0066).

The relative RSR did not decrease for all attacker types and there were no significant differences.

Combining the results, the acceptable k levels based on our criteria were $k=1$ for targeted attackers, $k=2$ for naive attackers, and at least $k=6$ for VPN attackers.

\section{Discussion} \label{sec:discussion}

Our %
results show that IP address truncation significantly affects the RBA risk score and reduces the probability of attack detection.
The truncation for VPN attackers resulted in a local maximum of the RSR at 12 bits, and thus apparently improved detection.
However, this was due to the fact that the VPN attacker only had an %
IP address range limited to %
the VPN service's server locations%
. Since the first IP address bits correspond to a node's geolocation, they were mostly distinct from legitimate users residing in different areas. 
Thus, truncating %
increased the risk scores for VPN attackers until 12 bit, as the probability for the global login history $p(FV^k)$ 
decreased but the one for the local history $p(FV^k | u, legit)$ remained constant. In contrast to that, targeted attackers also had a limited IP address range, but they were located in the same region as the legitimate users. Also, naive attackers had a large IP address range%
. Thus, in both cases, the differences between $p(FV^k)$ and $p(FV^k | u, legit)$ remained constant to similar levels until 12 bits.

Following that, and what our evaluation indicates, we do not recommend truncating more than three bits for a stable RBA performance in our case study scenario. %

K-anonymity increased the distinguishability between legitimate users and attackers, i.e., the RSR. This was due to the fact that this mechanism added new entries to the global login history. As a result, the overall probability for unknown feature values in the global login history $p(FV^k)$ decreased, making it harder for attackers to achieve a low risk score. However, this also decreased the detection of attackers, i.e., the TPR, in most cases, since $k$ more users had similar feature values in the data set. As a side effect of these results, unique feature values got less unique in total. Thus, due to the determined limit of $k=1$, k-anonymity for targeted attackers can only be achieved with degraded RBA performance.

The overhead produced by the additional entries increased with each $k$ (see Table~\ref{tab:k-anonymity-overhead}). It was even more than the data set itself at k\textgreater 3, which makes the current mechanism impractical for very large online services. To mitigate this issue, mechanisms could be introduced which remove some additional login entries when k-anonymity can be fulfilled after some time.

\begin{table}[]
	\centering
	\caption{Overhead created by additional login entries to achieve k-anonymity}
	\resizebox{0.6\linewidth}{!}{%
		\begin{tabular}{@{}lll@{}}
			\toprule
			k & Additional Entries & Increase to Baseline \\
			\midrule
			1 & 0 & 0.0 \\
			2 & 3928 & 0.41 \\
			3 & 7965 & 0.83 \\
			4 & 12013 & 1.26 \\
			5 & 16065 & 1.68 \\
			6 & 20120 & 2.11 \\
			\bottomrule
		\end{tabular}%
	}
	\label{tab:k-anonymity-overhead}
\end{table}

K-anonymity is not scalable with an increasing number of features~\cite{aggarwal_k-anonymity_2005}, while the other approaches are. Thus, sensible RBA privacy enhancements might be a combination of all outlined countermeasures, to ensure scalability.

Based on our results, we discuss privacy challenges and further research directions in the following.

\subsection{Privacy Challenges}

When integrating privacy into RBA systems, there are several challenges that should be considered in practice. We describe them below.

\inlineheading{Role of the IP Address Feature}
Using a combination of privacy enhancements for the IP address might be sufficient for some applications. However, %
this feature is still sensitive information. Thus, the question arises whether online services should consider privacy enhancing alternatives instead of storing the IP address%
. One alternative could be to derive only the region and ASN from the IP address, and discard the rest. 
Other approaches even enable identifying network anomalies, e.g., IP spoofing using a VPN connection, without having to rely on the IP address at all. For example, the server-originated round-trip time (RTT)~\cite{wiefling_whats_2021} can be used to estimate the distance between the user's device and the server location and may replace IP addresses as RBA features. 
As the RTTs vary based on the server location, they become useless for most re-identification attacks using leaked databases, as server locations are distributed in practice. They can even be enriched with random noise to further enhance privacy.

\inlineheading{Risk of Feature Stuffing}
Such considerations can be more and more important with widespread RBA adoption in the future. We assume that when databases with RBA feature values got stolen, this might have serious consequences for other services using RBA. In contrast to passwords, behavioral RBA feature values cannot be changed after compromise. Attackers can attempt to automatically reproduce these feature values on other websites. Thus, more privacy preserving alternatives that are hard to spoof for attackers might be crucial to mitigate largely scalable ``feature stuffing'' attacks.

\inlineheading{Handling Data Deletion Requests}
Further conflicts could arise with data protection regulations. Users are legally permitted to request data deletion. So when they request online services to delete their RBA feature data, they might lose RBA protection on their user accounts.

\subsection{Research Directions}

Our case study evaluation provided first insights on truncating feature values to increase privacy. As the results showed that this is possible to a certain degree while maintaining RBA performance, further work can investigate it for other types of features, e.g., the user agent string. %

The proposed k-anonymity mechanism can increase privacy regarding unique entries in the data set. However, users might still be identifiable when they have a combination of typical feature values, e.g., a home and a work IP address. This non-trivial task had been addressed in dynamic databases%
~\cite{garcia-alfaro_general_2018, xiao_m-invariance_2007}. Future work may investigate whether such mechanisms are also applicable to RBA.

As we could not reliably test the login history minimization approach with our data set, future work should investigate this on a medium to large-scale online service with regular use%
.

\section{Related Work} \label{sec:related-work}

Burkhard et al.~\cite{burkhart_risk-utility_2008} investigated truncating IP addresses in anomaly detection systems. They found that truncating more than four bits degraded the performance of these systems. Chew et al.~\cite{chew_privacy_2019} further evaluated IP truncation in intrusion detection systems. Their results showed that the detection accuracy in many of the tested classifiers decreased after removing more than 8 bits. %
Our study showed that three bits could be removed from the IP address to maintain RBA performance at the same time.

Both Safa et al.~\cite{cuppens-boulahia_privacy-preserving_2014}, and Blanco-Justicia and Domingo-Ferrer~\cite{blanco-justicia_efficient_2018} proposed privacy-preserving authentication models for implicit authentication using mobile devices. Their models relied on client-originated features, and the former also calculated risk scores on the client's device. However, this is not applicable to our RBA use case, as it relies on server-originated features and risk scores to prevent client-side spoofing.

To the best of our knowledge, there were no studies investigating privacy enhancements in RBA systems. However, some literature touched on privacy aspects related to RBA. Bonneau et al.~\cite{bonneau_privacy_2014} discussed privacy concerns of using additional features for authentication. They found that privacy preserving techniques might mitigate these concerns, but these had not been deployed in practice. We proposed and tested some techniques for the first time in our case study. Wiefling et al.~\cite{wiefling_more_2020} investigated RBA's usability and security perceptions. The results showed that users tended to reject providing phone numbers to online services for privacy reasons. They further studied RBA characteristics on a real-world online service~\cite{wiefling_whats_2021}, showing that the feature set can be very small to achieve good RBA performance. We demonstrated that the privacy can be further enhanced through different mechanisms.

\section{Conclusion} \label{sec:conclusion}

With a widespread use of RBA to protect users against attacks involving stolen credentials, more and more online services will potentially store sensitive feature data of their users, like IP addresses and browser identifiers, for long periods of time. Whenever such information is forwarded or leaked, it poses a potential threat to user privacy. To mitigate such threats, the design of RBA systems must balance security and privacy.

Our study results provide a first indication that RBA implementations used in current practice can be designed to become more privacy friendly. However, there are still challenges that have not been resolved in research to date. An important question is, e.g., how the IP address feature can be replaced with more privacy preserving alternatives. On the one hand, we assume that 
the IP address is very relevant for re-identification attacks~\cite{europol_sirius_2019}. Discarding it from the RBA login history can therefore increase privacy protection. On the other hand, the IP address is %
a feature providing strong security~\cite{wiefling_whats_2021}. Future research must carefully identify and analyze such trade-offs, so that RBA's user acceptance does not drop with the first data breach.

	\bibliographystyle{IEEEtranS}
	\bibliography{bibliography.bib}

\begin{thebibliography}{10}
\providecommand{\url}[1]{#1}
\csname url@samestyle\endcsname
\providecommand{\newblock}{\relax}
\providecommand{\bibinfo}[2]{#2}
\providecommand{\BIBentrySTDinterwordspacing}{\spaceskip=0pt\relax}
\providecommand{\BIBentryALTinterwordstretchfactor}{4}
\providecommand{\BIBentryALTinterwordspacing}{\spaceskip=\fontdimen2\font plus
\BIBentryALTinterwordstretchfactor\fontdimen3\font minus
  \fontdimen4\font\relax}
\providecommand{\BIBforeignlanguage}[2]{{%
\expandafter\ifx\csname l@#1\endcsname\relax
\typeout{** WARNING: IEEEtranS.bst: No hyphenation pattern has been}%
\typeout{** loaded for the language `#1'. Using the pattern for}%
\typeout{** the default language instead.}%
\else
\language=\csname l@#1\endcsname
\fi
#2}}
\providecommand{\BIBdecl}{\relax}
\BIBdecl

\bibitem{aggarwal_k-anonymity_2005}
C.~C. Aggarwal, ``On k-{Anonymity} and the {Curse} of {Dimensionality},'' in
  \emph{{VLDB} '05}.\hskip 1em plus 0.5em minus 0.4em\relax VLDB Endowment,
  Aug. 2005.

\bibitem{akamai_loyalty_2020}
{Akamai}, ``Loyalty for {Sale} – {Retail} and {Hospitality} {Fraud},''
  \emph{[state of the internet] / security}, vol.~6, no.~3, Oct. 2020.

\bibitem{almotairi_perception_2020}
K.~Almotairi and B.~Bataineh, ``Perception of {{Information Sensitivity}} for
  {{Internet Users}} in {{Saudi Arabia}},'' \emph{{AIP}}, vol.~9, no.~2, 2020.

\bibitem{blanco-justicia_efficient_2018}
A.~Blanco-Justicia and J.~Domingo-Ferrer, ``Efficient privacy-preserving
  implicit authentication,'' \emph{Computer Communications}, vol. 125, Jul.
  2018.

\bibitem{bonneau_privacy_2014}
J.~Bonneau, E.~W. Felten, P.~Mittal, and A.~Narayanan, ``Privacy concerns of
  implicit secondary factors for web authentication,'' in \emph{{WAY} '14},
  Jul. 2014.

\bibitem{burkhart_risk-utility_2008}
M.~Burkhart, D.~Brauckhoff, M.~May, and E.~Boschi, ``The risk-utility tradeoff
  for {IP} address truncation,'' in \emph{{NDA} '08}.\hskip 1em plus 0.5em
  minus 0.4em\relax ACM, 2008.

\bibitem{chew_privacy_2019}
Y.~J. Chew, S.~Y. Ooi, K.-S. Wong, and Y.~H. Pang, ``Privacy {Preserving} of
  {IP} {Address} through {Truncation} {Method} in {Network}-based {Intrusion}
  {Detection} {System},'' in \emph{{ICSCA} '19}.\hskip 1em plus 0.5em minus
  0.4em\relax ACM, 2019.

\bibitem{european_union_gdpr_2016}
{European Union}, ``General {Data} {Protection} {Regulation},'' May 2016,
  {Regulation} (EU) 2016/679.

\bibitem{europol_sirius_2019}
{Europol}, ``{SIRIUS} {EU} {Digital} {Evidence} {Situation} {Report} 2019,''
  Dec. 2019.

\bibitem{federal_committee_on_statistical_methodology_report_2005}
{Federal Committee on Statistical Methodology}, ``Report on {Statistical}
  {Disclosure},'' Dec. 2005.

\bibitem{firehol_all_2020}
\BIBentryALTinterwordspacing
{FireHOL}, ``All cybercrime ip feeds,'' Aug. 2020. [Online]. Available:
  \url{http://iplists.firehol.org/?ipset=firehol\_level4}
\BIBentrySTDinterwordspacing

\bibitem{freeman_who_2016}
D.~Freeman, S.~Jain, M.~D\"urmuth, B.~Biggio, and G.~Giacinto, ``Who {Are}
  {You}? {A} {Statistical} {Approach} to {Measuring} {User} {Authenticity},''
  in \emph{{NDSS} '16}.\hskip 1em plus 0.5em minus 0.4em\relax Internet
  Society, Feb. 2016.

\bibitem{grassi_digital_2017}
P.~A. Grassi \emph{et~al.}, ``Digital identity guidelines: authentication and
  lifecycle management,'' National Institute of Standards and Technology, Tech.
  Rep. NIST SP 800-63b, Jun. 2017.

\bibitem{haber_attack_2020}
M.~J. Haber, ``Attack {Vectors},'' in \emph{Privileged {Attack} {Vectors}:
  {Building} {Effective} {Cyber}-{Defense} {Strategies} to {Protect}
  {Organizations}}.\hskip 1em plus 0.5em minus 0.4em\relax Apress, 2020.

\bibitem{hurkala_architecture_2014}
A.~Hurkała and J.~Hurkała, ``Architecture of context-risk-aware
  authentication system for web environments,'' in \emph{{ICIEIS} '14}, 2014.

\bibitem{iso_isoiec_2011}
ISO, \emph{\BIBforeignlanguage{en}{{{ISO}}/{{IEC}} 29100:2011({{E}}):
  {{Information}} Technology \textemdash{} {{Security}} Techniques
  \textemdash{} {{Privacy}} Framework}}.\hskip 1em plus 0.5em minus 0.4em\relax
  {ISO/IEC}, 2011.

\bibitem{llewellyn-jones_cracking_2017}
D.~Llewellyn-Jones and G.~Rymer, ``Cracking {PwdHash}: {A} {Bruteforce}
  {Attack} on {Client}-side {Password} {Hashing}.''\hskip 1em plus 0.5em minus
  0.4em\relax Apollo, 2017.

\bibitem{markos_new_2018}
E.~Markos, L.~I. Labrecque, and G.~R. Milne, ``A {{New Information Lens}}:
  {{The Self}}-concept and {{Exchange Context}} as a {{Means}} to {{Understand
  Information Sensitivity}} of {{Anonymous}} and {{Personal Identifying
  Information}},'' \emph{{JIM}}, vol.~42, 2018.

\bibitem{markos_information_2017}
E.~Markos, G.~R. Milne, and J.~W. Peltier, ``Information {{Sensitivity}} and
  {{Willingness}} to {{Provide Continua}}: {{A Comparative Privacy Study}} of
  the {{United States}} and {{Brazil}},'' \emph{{JPP\&M}}, 2017.

\bibitem{mccallister_guide_2010}
E.~McCallister, T.~Grance, and K.~A. Scarfone, ``Guide to protecting the
  confidentiality of {Personally} {Identifiable} {Information} ({PII}),'' Tech.
  Rep. NIST SP 800-122, 2010.

\bibitem{molloy_risk-based_2012}
I.~Molloy, L.~Dickens, C.~Morisset, P.-C. Cheng, J.~Lobo, and A.~Russo,
  ``Risk-based {Security} {Decisions} {Under} {Uncertainty},'' in
  \emph{{CODASPY} '12}.\hskip 1em plus 0.5em minus 0.4em\relax ACM, Feb. 2012.

\bibitem{moriarty_pkcs_2017}
K.~Moriarty, B.~Kaliski, and A.~Rusch, ``Pkcs \#5: Password-based cryptography
  specification version 2.1,'' RFC 8018, January 2017.

\bibitem{national_cyber_security_centre_cloud_2018}
{National Cyber Security Centre}, ``Cloud security guidance: 10, {Identity} and
  authentication,'' Tech. Rep., Nov. 2018.

\bibitem{pugliese_long-term_2020}
G.~Pugliese, C.~Riess, F.~Gassmann, and Z.~Benenson, ``Long-{Term}
  {Observation} on {Browser} {Fingerprinting}: {Users}’ {Trackability} and
  {Perspective},'' \emph{{PoPETS}}, vol. 2020, no.~2, Apr. 2020.

\bibitem{quermann_state_2018}
N.~Quermann, M.~Harbach, and M.~Dürmuth, ``The {State} of {User}
  {Authentication} in the {Wild},'' in \emph{{WAY} '18}, Aug. 2018.

\bibitem{cuppens-boulahia_privacy-preserving_2014}
N.~A. Safa, R.~Safavi-Naini, and S.~F. Shahandashti, ``Privacy-{Preserving}
  {Implicit} {Authentication},'' in \emph{{IFIP} {SEC} '14}.\hskip 1em plus
  0.5em minus 0.4em\relax Springer, 2014.

\bibitem{garcia-alfaro_general_2018}
J.~Salas and V.~Torra, ``A {General} {Algorithm} for k-anonymity on {Dynamic}
  {Databases},'' in \emph{{DPM} '18}.\hskip 1em plus 0.5em minus 0.4em\relax
  Springer, 2018.

\bibitem{schomakers_internet_2019}
E.-M. Schomakers, C.~Lidynia, D.~Müllmann, and M.~Ziefle, ``Internet users’
  perceptions of information sensitivity – insights from {Germany},''
  \emph{{IJIM}}, vol.~46, Jun. 2019.

\bibitem{schomakers_all_2020}
E.-M. Schomakers, C.~Lidynia, and M.~Ziefle, ``All of me? {{Users}}'
  preferences for privacy-preserving data markets and the importance of
  anonymity,'' \emph{Electronic Markets}, vol.~30, no.~3, Feb. 2020.

\bibitem{california_ccpa_2018}
{State of California}, ``{California} {Consumer} {Privacy} {Act},'' Jun. 2018,
  {Assembly} {Bill} {No.} 375.

\bibitem{steinegger_risk-based_2016}
R.~H. Steinegger, D.~Deckers, P.~Giessler, and S.~Abeck,
  ``\BIBforeignlanguage{en}{Risk-based authenticator for web applications},''
  in \emph{\BIBforeignlanguage{en}{{EuroPlop} '16}}.\hskip 1em plus 0.5em minus
  0.4em\relax ACM, Jun. 2016.

\bibitem{sweeney_k-anonymity_2002}
L.~Sweeney, ``k-anonymity: {A} model for protecting privacy,'' \emph{{IJUFKS}},
  vol.~10, no.~05, Oct. 2002.

\bibitem{venkatadri_investigating_2018}
G.~Venkatadri, E.~Lucherini, P.~Sapiezynski, and A.~Mislove, ``Investigating
  sources of {PII} used in {Facebook}’s targeted advertising,''
  \emph{{PoPETS}}, vol. 2019, Jan. 2019.

\bibitem{wiefling_whats_2021}
S.~Wiefling, M.~D\"{u}rmuth, and L.~Lo~Iacono, ``What’s in {Score} for
  {Website} {Users}: {A} {Data}-driven {Long}-term {Study} on {Risk}-based
  {Authentication} {Characteristics},'' in \emph{{FC} '21}.\hskip 1em plus
  0.5em minus 0.4em\relax Springer, Mar. 2021.

\bibitem{wiefling_more_2020}
S.~Wiefling, M.~Dürmuth, and L.~Lo~Iacono, ``More {Than} {Just} {Good}
  {Passwords}? {A} {Study} on {Usability} and {Security} {Perceptions} of
  {Risk}-based {Authentication},'' in \emph{{ACSAC} '20}.\hskip 1em plus 0.5em
  minus 0.4em\relax ACM, Dec. 2020.

\bibitem{wiefling_is_2019}
S.~Wiefling, L.~Lo~Iacono, and M.~D\"urmuth, ``Is {This} {Really} {You}? {An}
  {Empirical} {Study} on {Risk}-{Based} {Authentication} {Applied} in the
  {Wild},'' in \emph{{IFIP} {SEC} '19}.\hskip 1em plus 0.5em minus 0.4em\relax
  Springer, Jun. 2019.

\bibitem{xiao_m-invariance_2007}
X.~Xiao and Y.~Tao, ``M-invariance: towards privacy preserving re-publication
  of dynamic datasets,'' in \emph{{SIGMOD} '07}.\hskip 1em plus 0.5em minus
  0.4em\relax ACM, 2007.

\end{thebibliography}
	
\end{document}